\documentclass[amsmath,amssymb,aps,twocolumn]{revtex4-2}

\usepackage{amssymb}
\usepackage{natbib}
\usepackage{graphicx}
\usepackage{amsmath}
\usepackage[bookmarks = false]{hyperref}
\usepackage{bm}
\usepackage[all]{hypcap}
\usepackage{graphicx}
\usepackage{colortbl}
\usepackage{booktabs}
\usepackage{enumerate}
\usepackage{enumitem}
\usepackage{braket}
\usepackage{mathrsfs}
\usepackage{multirow}
\usepackage{upgreek}
\usepackage{textcomp}

\usepackage{lineno}

\usepackage{array}

\renewcommand{\section}[1]{\vspace{\baselineskip}{\centering \textbf{#1}\\}\vspace{0.5\baselineskip}}

\begin{document}

\title{Single-shot measurement of a Rydberg superatom via collective photon burst}
\author{Chao-Wei Yang$^{1,\,2}$}
\author{Jun Li$^{1,\,2}$}
\author{Ming-Ti Zhou$^{1,\,2}$}
\author{Xiao Jiang$^{1,\,2}$}
\author{Xiao-Hui Bao$^{1,\,2}$}
\author{Jian-Wei Pan$^{1,\,2}$}

\affiliation{$^1$Hefei National Laboratory for Physical Sciences at Microscale and Department
of Modern Physics, University of Science and Technology of China, Hefei,
Anhui 230026, China}
\affiliation{$^2$CAS Center for Excellence and Synergetic Innovation Center in Quantum
Information and Quantum Physics, University of Science and Technology of
China, Hefei, Anhui 230026, China}

\begin{abstract}
With Rydberg dipole interactions, a mesoscopic atomic ensemble may behave like a two-level single atom, resulting in the so-called picture of superatom. It is in potential a strong candidate as a qubit in quantum information science, especially for efficient coupling with single photons via collective enhancement that is essential for building quantum internet to connect remote quantum computers. Previously, preliminary studies have been carried out in demonstrating basic concept of Rydberg superatom, a single-photon source, and entanglement with a single photon, etc. While a crucial element of single-shot qubit measurement is still missing. Here we realize the deterministic measurement of a superatom qubit via photon burst in a single shot. We make use of a low-finesse ring cavity to enhance the atom-photon interaction and obtain an in-fiber retrieval efficiency of 44\%. Harnessing dipole interaction between two Rydberg levels, we may either create a sequence of multiple single photons or nothing, conditioned on the initial qubit state. We achieve a single-shot measurement fidelity of 93.2\% in 4.8 $\upmu$s. Our work complements the experimental toolbox of harnessing Rydberg superatom for quantum information applications.
\end{abstract}

\maketitle

Quantum internet~\cite{kimble2008quantum} aims to connect remote quantum nodes efficiently that enables a number of  profound applications~\cite{wehner2018quantum}, such as distributed quantum computing, multi-party quantum communication and quantum repeater. The physical realization of quantum internet requires a system that can couple with single photons efficiently. An ensemble of atoms very suits for this purpose~\cite{Sangouard2011}, due to the collectively enhanced atom-photon interaction. A qubit in an atomic ensemble is usually measured via converting to a single-photon field and performing photon counting. This approach has limited overall efficiency, due to inefficiencies in retrieval, transmission and detection. When photon detection fails, no measurement result can be given, making a commonly single-shot measurement impossible. State-dependent fluorescence detection is ubiquitously used for detecting single-atom qubits~\cite{gibbons_nondestructive_2011,fuhrmanek_free-space_2011,kwon_parallel_2017}, while its adoption for an atomic ensemble qubit is very challenging due to off-scattering of reservoir atoms. An effective way is pushing off the reservoir atoms~\cite{ebert_coherence_2015} and applying traditional single-atom fluorescence detection. Nevertheless, such an approach is destructive and requires atom reloading for subsequent experimental trials. Similar to single atoms, a strongly coupled cavity-QED system~\cite{chen_all-optical_2013} also provides a route for lossless detection of collective atomic excitations, albeit requiring a technically demanding setup.

Rydberg interactions~\cite{Saffman2010} in an atomic ensemble enable the realization of a vast range of new quantum optical phenomena, such as single-photon interactions ~\cite{Peyronel2012a,Firstenberg2013a,thompson_symmetry-protected_2017,cantu_repulsive_2020}, transistor~\cite{tiarks_single-photon_2014,gorniaczyk_single-photon_2014,gorniaczyk_enhancement_2016}, contactless nonlinear optics~\cite{busche_contactless_2017}, and photon-photon gate~\cite{tiarks_photonphoton_2019}. If the size is small enough (typically $< 10$ $\upmu$m), an ensemble of atoms can behave like a two-level atom~\cite{Saffman2010}. Such a superatom has many benefits. In comparison with single atoms, the interaction with single photons are collectively enhanced. In comparison with traditional ensemble approach, it has the benefit of deterministic entanglement preparation. Besides, the long-range dipole interaction also provides a promising interface~\cite{Saffman2005,petrosyan_deterministic_2018,grankin_free-space_2018} for quantum computers with single atoms in arrayed optical tweezers~\cite{bluvstein_controlling_2021}, and the interaction with microwave photons also enables a promising interface~\cite{petrosyan_microwave_2019} for super-conducting quantum computers~\cite{arute_quantum_2019,gong_quantum_2021}. Significant experimental progresses have been achieved so far, such as the observation collective Rabi oscillation~\cite{Dudin2012e}, on demand single-photon source~\cite{Dudin2012sc} and atom-photon entanglement~\cite{Li2013}. It was also reported of creating multiple superatoms and interfering them for entanglement production~\cite{Li2016,Li2019}. With these developments, targeting advanced quantum applications (e.g. constructing a quantum network of multiple superatoms), it becomes more and more urgent to extend the experimental toolbox of superatom manipulation, such as improving the single-photon retrieval efficiency and developing the technology of deterministic single-shot measurement. In previous studies of single-photon transistor~\cite{tiarks_single-photon_2014,gorniaczyk_single-photon_2014,gorniaczyk_enhancement_2016}, a single photon can control the transmission of many photons with an achievable gain as high as 100, and probabilistic collective excitations were measured. Nevertheless, these realizations have not reached the superatom regime and did require additional experimental setups.

\begin{figure*}[htb]
	\centering
	\includegraphics[width= 0.8\textwidth]{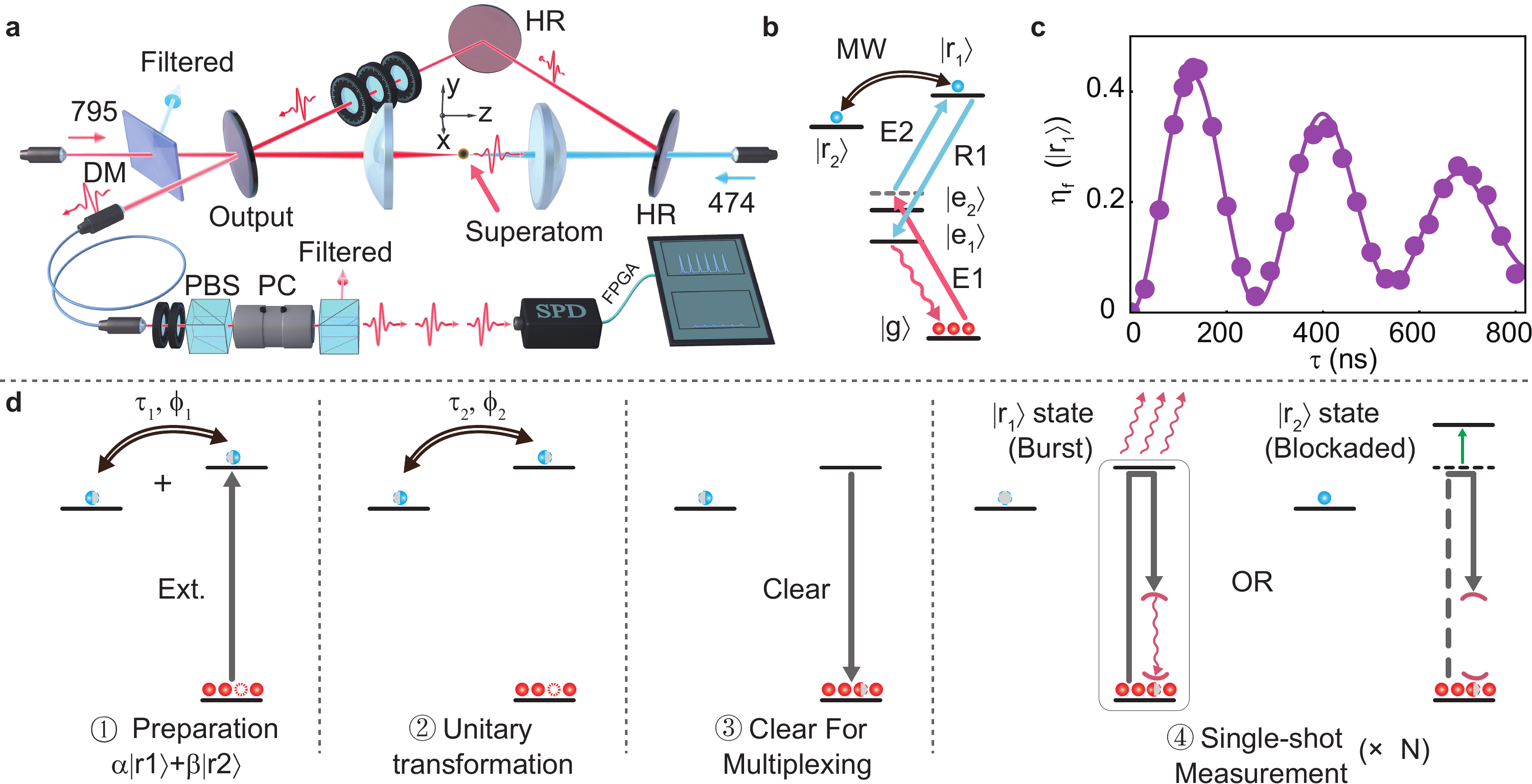}
	\caption{\textbf{Schematic of the cavity enhanced Rydberg superatom and single-shot measurement.} (a) Experimental setup. A tiny $^{87}$Rb ensemble is confined in an optical trap, with a freespace ring cavity placed around. A 795 nm (E1) and a reversely collinear 474 nm (E2/R1) laser couple with the superatom, realizing two-photon excitation and retrieving as shown in (b). And the cavity mirrors are coated to only form a cavity at 795 nm, but allowing the 474 nm beam fully pass. The single photon is retrieved along the anticlockwise cavity mode. A pockels cell (PC) is inserted to realize dynamical noise filtering (see Supplemental Material), and the SPD is coupled with single mode fiber (SMF, not shown). A DM (dichroic mirror) is used to seperate 474 and 795 nm lasers. (b) Energy level scheme. After two-photon excitation, A Rydberg superatom ($\ket{r_1}=\ket{91S_{1/2}, m_j=1/2}$) can be created with the excitation area within the blockade radius. Then the $\ket{r_1}$ state can be transferred to another state $\ket{r_2}=\ket{91S_{1/2}, m_j=-1/2}$ with MW Raman pulses. (c) Collective Rabi oscillation between $\ket{g}$ and $\ket{r_1}$. The single photon is directly retrieved with different excitation period. The peak in-fiber efficiency $\eta_f$ with an optical depth (OD) of 1.9, is measured to be 44\%. $\eta_f$ refers to the efficiency with the single photon collected into SMF. (d) Experimental procedures. After superatom qubit preparation, the projection measurements are performed with another MW Raman pulse and photon burst, by repeating excitation and retrieving for several times.}
	\label{fig:setup}
\end{figure*}

\begin{figure}[htp]
	\centering
	\includegraphics[width=.9\columnwidth]{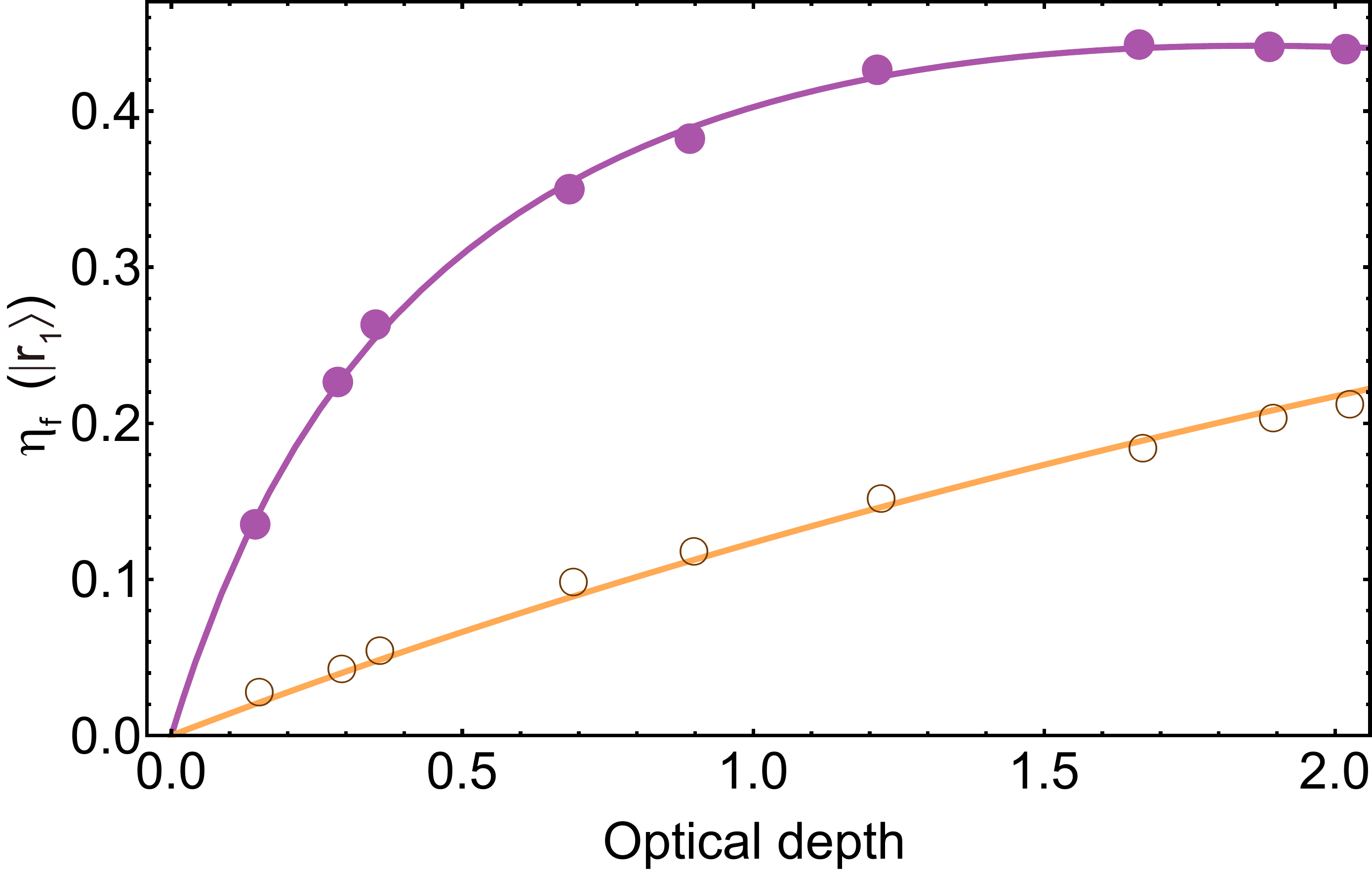}
	\caption{\textbf{Performance of the cavity enhancement with different OD.} In freespace, $\eta_f$ almost linearly increases with OD (Orange, hollow circle), with the maximum efficiency reaching 21\%. In caivty, $\eta_f$ shows saturation with large OD (Purple, solid circle), with the maximum efficiency reaching 44\%. }
	\label{fig:CaviEnhan}
\end{figure}

In this article, we report the experimental realization of single-shot measurement for a Rydberg superatom qubit via photon burst. Our scheme is shown in Fig.~\ref{fig:setup}d. We make use of Rydberg blockade~\cite{Saffman2010} between $|r_1\rangle$ and $|r_2\rangle$ to probe the population in $|r_2\rangle$ via controlled multi-photon generation. If there is no population in $|r_2\rangle$, we can repeat the process of excitation and retrieving for $|r_1\rangle$ to create a sequence of single photons. While if $|r_2\rangle$ is populated, this process will not generate any photon. Therefore, by counting the number of photons, we can measure the population in $|r_2\rangle$. Furthermore, by applying a microwave (MW) field that couples the transition $|r_1\rangle \leftrightarrow |r_2\rangle$, we can prepare an arbitrary single qubit state and perform measurements in arbitrary bases. This scheme simply makes use of the build-in single-photon interface for qubit measurement, which is advantageous experimentally. Then in order to obtain a high fidelity in a short measurement duration, it is crucially important to improve photon detection efficiency. By making use of a low-finesse cavity to further enhance the collective atom-photon interaction, we achieve a high in-fiber retrieval efficiency $\eta_f$ = 44\%. By repeating the exciting and retrieving process for 12 times in 4.8 $\upmu$s, we detect 2.63 photons for a superatom in $|r_1\rangle$ and 0.19 photons for a superatom in $|r_2\rangle$, which gives a single-shot measurement fidelity of 91.3\%. After correcting the preparation infidelity, the single-shot fidelity reaches 93.2\%. Then for a qubit in the superposition of $|r_1\rangle$ and $|r_2\rangle$, we achieve a close tomography fidelity.

\begin{figure*}[htb]
	\centering
	\includegraphics[width=.75\textwidth]{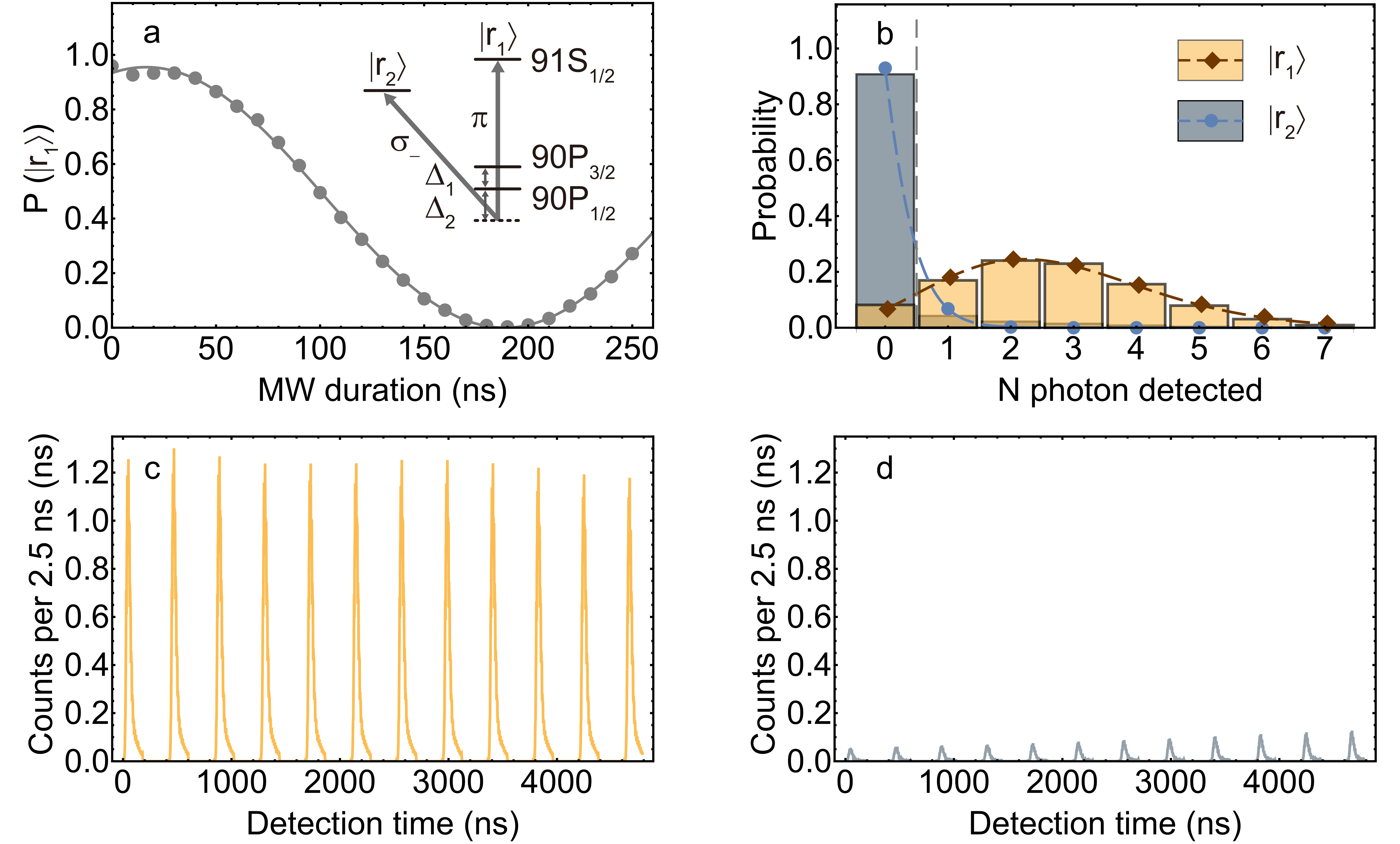}
	\caption{\textbf{Single-shot measurement of the $\ket{r_1}$ and $\ket{r_2}$ state.} (a) MW Raman driven Rabi oscillation between $\ket{r_1}$ and $\ket{r_2}$. As shown in the dip, $\ket{r_1}$ can be transferred to $\ket{r_2}$ with 99.7$\%$ fidelity. $P_{\ket{r_1}}$ corresponds to the population of $\ket{r_1}$ by detecting the retrieved signals. The primary MW Raman coupling channels is shown in the inset. $\Delta_2$=390 MHz. $\Delta_1$=130 MHz. (b) Photon number statistical distribution without correcting the preparation infidelity of detecting N photons in 4.8 $\upmu$s. This is averaged from $3.6\times10^6$ cycles in 560 s (including the loading period). The dashed gray line seperates two regions of detecting 0 and $\geqslant$ 1 photons. The histograms are from the data, and fitted with Poisson distribution shown as dots and line. (c) and (d) are the measured temporal profiles with the superatom prepared in $\ket{r_1}$ and $\ket{r_2}$ respectively. Photon counts are registered with a bin size of 2.5 ns. In comparison, significant photon suppression is observed in (d). Residual signal is mainly coming from the infidelity of superatom preparation and gradual Rydberg atom loss during measurement.}
	\label{fig:EnhanMeas}
\end{figure*}

Our detailed experimental setup is sketched in Fig.~\ref{fig:setup}. In order to reach the superatom regime, we capture a tiny atomic ensemble with the thickness of $\sim$3 $\upmu$m, using a 852 nm optical trap. Then we deterministically prepare a Rydberg single excitation making use of two-photon excitation with the E1 and E2 pulses shown in Fig.~\ref{fig:setup}b, to restrict the $1/e^2$ radius of excitation area within 6.5 $\upmu\rm{m}$. Then we observe the collective Rabi oscillation as shown in Fig.~\ref{fig:setup}c with the collective Rabi frequency of 2$\pi\times$2.9 MHz. The blockade radius is estimated to be $\sim$ 9.0 $\upmu$m, which is much larger than the excitation area. By setting the $\pi$ pulse excitation period, we realize the deterministic superatom preparation. We choose firstly preparing the superatom into the qubit state of $\ket{r_1}=\ket{91S_{1/2}, m_j=1/2}$ and $\ket{r_2}=\ket{91S_{1/2}, m_j=-1/2}$. With two independent MW pulses, we realize high fidelity MW Raman operation between $\ket{r_1}$ and $\ket{r_2}$ state, and prepare them into arbitrary qubit state by setting different pulse areas and phases in the 1st operation of Fig.~\ref{fig:setup}d. Then in order to perform single-shot measurement of the qubit state, we make use of the Rydberg interaction between $\ket{r_1}$ and $\ket{r_2}$, and collective photon burst with the Rydberg superatom combined with an optical cavity as shown in Fig.~\ref{fig:setup}a. Benefitting from the deterministic preparation, we can clear $\ket{r_1}$ state, then the population of $\ket{r_2}$ will identify the qubit state. Afterwards the cleared $\ket{r_1}$ channel will be multiplexed to implement collective photon burst.

The high efficiency photon-superatom interface is of great significance in quantum internet and our single-shot measurement scheme. Due to the limited optical depth (OD) of the superatom, we set up a ring cavity~\cite{bao2012efficient,Yang2016} around the superatom to enhance the superatom-photon coupling strength, to realize efficiently emission and detection of photons in a short period which is defined as collective photon burst. Until now, although optical cavity has been used to achieve strong coupling with Rydberg atoms~\cite{jia2018strongly}, performing high efficiency photon collection and detection is still missing. In our platform, the caivty is built with three customized planar mirrors, and two aspherical lens to create the cavity waist radius of 6.5 $\upmu$m. The left mirror in Fig.~\ref{fig:setup}a is the output mirror with 78$\%$ reflectivity. Due to the intra-cavity loss of $\sim$7\%, we choose to use higher output proportion to make sure of high output efficiency of 80\%, with the other two mirrors designed as high-reflectivity mirrors. The cavity has a finesse of $\mathscr{F}$ = 19.5, realizing intrinsic enhancement of $2\mathscr{F}/\pi$=12.4~\cite{simon2010cavity,bao2012efficient} of the cooperativity, and then the photons leaving the cavity are collected into the single mode fiber (SMF) with the efficiency of 85.9$\%$, and detected with a SPD with the efficiency $\eta_{SPD}$ of 68\%.

The retrieval efficiencies with and without cavity enhancement are shown in Fig.~\ref{fig:CaviEnhan}. We test different results by changing the OD. At each point, we prepare a superatom, and detect the retrieved single photons. There we measure the single-photon retrieval efficiency $\eta$ with the SPD, then we get the in-fiber efficiency $\eta_f=\eta/\eta_{SPD}$.  As shown in Fig.~\ref{fig:CaviEnhan}, the efficiency increases linearly with OD in freespace, but in cavity it gradually shows saturation with large OD. And we get the highest $\eta_f$ of 21$\%$ in freespace (with collection efficiency of 90\%), and 44\% in cavity. So we get the intrinsic retrieval efficiencies of 24\% in freespace and 64\% in cavity with OD=1.9. After fitting the intrinsic retrieval efficiencies with the $k\cdot OD/(k\cdot OD+1)$ model (see Supplemental Material), we find the result in cavity seems to be 17.7\% lower than the theoretical prediction. We infer that the main reason comes from the limited size of the Read beam, which will cause dephasing because of the inhomogeneous AC stark shift. \cite{simon2010cavity}

\begin{figure}[htb]	
	\centering
	\includegraphics[width=\columnwidth]{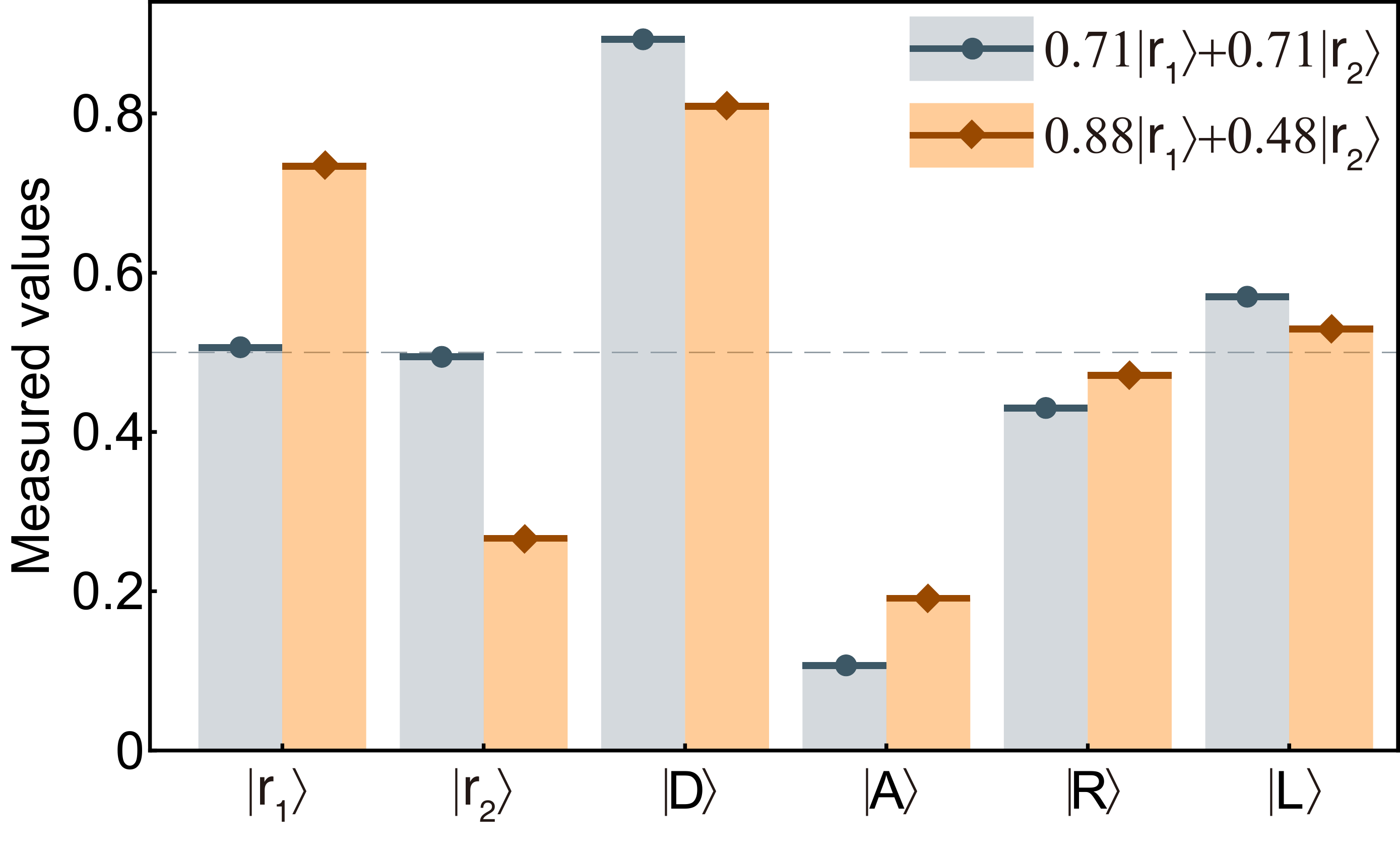}
	\caption{\textbf{Projection measurement and tomography with single-shot measurement.} The projection measurement is implemented 40 ns after initial state preparation, with the pulse period and relative phase rapidly switched to measure in different bases. $\ket{D}$ = $(\ket{r_1} + \ket{r_2})/\sqrt{2}$, $\ket{A}$ = $(\ket{r_1} - \ket{r_2})/\sqrt{2}$, $\ket{R}$ = $(\ket{r_1} + i \ket{r_2})/\sqrt{2}$, and $\ket{L}$ = $(\ket{r_1} - i \ket{r_2})/\sqrt{2}$. The fidelity is calculated to be 89.3$\%$ for $0.71 \ket{r_1} + 0.71 \ket{r_2}$ and 88.6$\%$ for $0.88 \ket{r_1} + 0.48 \ket{r_2}$ by rebuilding the density matrix.}
	\label{fig:tomography}
\end{figure}

With MW operations and the high efficiency collective photon burst, we prepare the qubit respectively in $\ket{r_1}$ and $\ket{r_2}$, and performing single-shot measurements. The main results are shown in Fig.~\ref{fig:EnhanMeas}. In order to measure $\ket{r_1}$, we make use of photon burst, by repeating excitation and retrieving of 12 times, and detecting 2.63 photons in 4.8 $\upmu$s with the photons profile in temporal domain shown in Fig.~\ref{fig:EnhanMeas}c. Similarly we prepare the superatom into $\ket{r_2}$ but with the MW $\pi$ pulse as shown in Fig.~\ref{fig:EnhanMeas}a, with the transfer fidelity of 99.7\%. Then we get the contrastive result in Fig.~\ref{fig:EnhanMeas}d, in which the photon emission is intensely suppressed to 0.19 photons by $\ket{r_2}$. Furthermore, except the mean photon numbers, we also analyse the statistical distributions of detecting different photon numbers, as shown in Fig.~\ref{fig:EnhanMeas}b. As we can see, for the $\ket{r_1}$ state it has the intrinsic noise of Poisson distribution (zero photon percentage of $e^{-N_{photon}}$), thus there is 8.2$\%$ zero photon events. As for the $\ket{r_2}$ state, although it is almost fully suppressed to zero, we still have tiny chance to detect one photon. The reason is from the dark count noise of 1.2$\%$ due to residual excitation noise. If divided by $N$ = 0 photon and  $N \geqslant$ 1 photon (dashed vertical line), the probability of detecting the superatom in $\ket{r_i}$ when prepared in $\ket{r_i}$ is calculated to be 90.8\% ($\ket{r_2}$) and 91.8\% ($\ket{r_1}$), with the average value of 91.3\% defined as the raw fidelity of single-shot measurement. After correcting the preparation infidelity of 4.5\% (see Supplemental Material), the photon emission of $\ket{r_2}$ is calculated to be actually suppressed to 0.08 photons, and the corrected fidelity is calculated to be 93.2\% with the respective probability of 94.6\% ($\ket{r_2}$) and 91.8\% ($\ket{r_1}$). Better fidelity can be achieved by harnessing a superconducting single-photon detector with much higher efficiency, and confining atom movement in Rydberg state~\cite{Li2013,Lampen2018,barredo2020three} with optical dipole trap etc.

Finally, we prepare the superatom into the superposition states of $\alpha\ket{r_1}+\beta\ket{r_2}$. With unitary transformation and single-shot measurement, we can demonstrate the qubit state tomography by perform projective measurements in arbitrary bases. Firstly, the superposition states $(\ket{r_1}+\ket{r_2})/\sqrt{2}$ is prepared with a $\pi/2$ MW pulse. Then we bring in another independent MW pulse with different pulse durations and relative phases to perform arbitrary unitary transformation. Without the second pulse, we can directly measure in the eigenstate bases of $\ket{r_2}$ or $\ket{r_1}$. Then with the $\pi$/2 pulse and different relative phase of 0 and $\pi$/2, we can measure in (anti)diagonal and circular bases of $(\ket{r_1}\pm\ket{r_2})/\sqrt{2}$ and $(\ket{r_1}\pm i\ket{r_2})/\sqrt{2}$ The projective measurements results are shown in Fig.~\ref{fig:tomography}. By calculating the Stokes parameters and rebuilding the density matrix~\cite{altepeter2005photonic}, we get a fidelity of 89.3$\%$ for $(\ket{r_1}+\ket{r_2})/\sqrt{2}$ state. Similarly, for another initial state $0.88 \ket{r_1} + 0.47 \ket{r_2}$, we get a fidelity of 88.6\%. We infer that the slight decrease of the fidelity compared with Fig.~\ref{fig:EnhanMeas}b mainly comes from the infidelity of the $\pi/2$ MW pulse, limited by the mechanical delay line which is used here to adjust the relative phase, which has low precision, and unstable insertion loss when adjusting the phases. It can be replaced by electronic elements in the future.

To summarize, we have realized the single-shot measurement of a Rydberg superatom qubit by making use of its efficient single-photon interface. Blockade between two Rydberg levels enables the controlled generation of sequential single photons. MW Raman coupling further enables the ability of measuring in a super-positional basis. And the enhancement via a low-finesse cavity boosts the single-photon retrieval efficiency significantly, which will not only benefit entangling rate between remote quantum nodes, but also plays an essential role in achieving high-fidelity measurement in a short duration. The technologies developed will enable a series of new capabilities that are not possible for traditional large atomic ensembles, such as performing device-independent quantum key distribution, deterministic entanglement swapping~\cite{pompili_realization_2021} for quantum repeater and quantum networking.

This work was supported by National Key R\&D Program of China (No. 2017YFA0303902), Anhui Initiative in Quantum Information Technologies, National Natural Science Foundation of China, and the Chinese Academy of Sciences.

\bibliography{myref}

\section{Supplemental Material}

\textbf{General information.}\ \
Our Rydberg ensemble is generated by firstly preparing a $^{87}$Rb cold atom ensemble with a magneto-optical trap (MOT), and doppler cooled to about 20 $\upmu$K. Then an 852 nm far-detuning single-beam optical trap is used to load the atoms from the ensemble. The 852 nm trap beam is shaped to an elliptical gaussian beam with a waist radius of 39 $\upmu\rm{m} \times $ 8.3 $\upmu$m propagating along the horizontal direction and orthogonal to the quantum axis. The maximum power can reach 710 mW and provides a maximum potential of 640 $\upmu$K. The finally captured atom ensemble has a $1/e^2$ waist radius of 7.15 $\upmu\rm{m} \times $70 $\upmu\rm{m} \times$1.43 $\upmu$m, and has a density on the order of $10^{12}$ atoms/cm$^3$. Together with the excitation area selection of $\sim$ 6.5 $\upmu$m radius, we achieve a tiny Rydberg blockade area within 6.5 $\upmu\rm{m} \times $6.5 $\upmu\rm{m} \times$1.43 $\upmu$m radius.

For each sequence, the loading period costs 100 ms, then the experiment cycles repeat 1000 times. In each cycle, the trap beam is pulsed turned off to perform other operations. And the locking cavity beam is also pulsed switched during molasses and operation period, and keeps locking during 30 ms operation time,  with the frequency set close to repumper laser to avoid severe atom blowing.

In the photon burst, a pockels cell is inserted to realize dynamical noise filtering, with additional loss of 15\%. Because of the compact time sequence in photon burst, the excitation pulse of 795 nm will directly enter the SPD $\sim$ 50 ns before the single photons, which will cause intense after-pulse noise. Thus we use a pockels cell to rapidly switch the pulse to the orthogonal polarization, then the second PBS will filter the noise in the temporal domain (Fig.~\ref{fig:setup}a). The first PBS is used to improve the extinction ratio, and the single photons can directly pass by carefully adjusting the incident polarization.

\textbf{Cavity enhancement.}\ \
The cavity has a finesse of 19.5 with the internal transmission of the cavity measured to be 96.5\% mainly due to coating loss. And we infer that the additional loss of 3.5\% comes from the cavity mode mismatch, due to the performance of two aspherical lens of EFI = 25 mm, which are used to create a small cavity mode waist. But their performance also seems to cause some distortion, and limits the quality of the cavity. The fiber collection efficiency of 85.9\% is calculated by directly measuring the transmitted weak probe beam from the cavity. Comparing and fitting the results in free space and in cavity, we find it basically agrees with the theory (fitted with $p\cdot C/(C+1)$, with $C=k\cdot OD$ in freespace and $C=k\cdot OD\cdot 2\mathscr{F}/\pi$ in cavity, k and p is the fitting parameter)\cite{simon2010cavity,bao2012efficient}. But the curve in cavity saturates faster than theory prediction as discussed in the main text. We infer it come from the limited size of the Read beam. Besides, we also test the situation with $\omega_{474}$ = 10.5 $\upmu$m, and it shows much more obvious saturation and decrease with large OD (not shown).

\textbf{Superatom preparation efficiency.}\ \
We compare the first peaks in Fig.~\ref{fig:EnhanMeas}(c) and (d), and make use of their ratio to estimate superatom preparation efficiency. The first peak in Fig.~\ref{fig:EnhanMeas}(c) shows the average photon brightness of $|r_1\rangle$ retrieval. Ideal superatom preparation in $|r_2\rangle$ with unity efficiency will in principle fully blockades photon retrieval through $|r_1\rangle$. While imperfect superatom preparation will lead to the emerging of tiny peaks shown in Fig.~\ref{fig:EnhanMeas}(d), as some experimental trials fail in a $|r_2\rangle$ superatom creation and have no influence over $|r_1\rangle$ retrieval. In this way, we estimate the superatom preparation efficiency to be 95.5\%.

\textbf{Microwave operation.}\ \
The qubit preparation is implemented with MW Raman pulses between $\ket{r_1}$ and $\ket{r_2}$. Different from $nP_J$ states, they have symmetrical blockade and will not cause the blockade vanishing in some directions. The Raman operation is formed with two horn antennas of horizontal (H) and circular ($\sigma_-$) polarization which are about 0.5 m away from the atoms. The purity of polarization is not so good because of the metal components around the atoms which will influence the transmission of the MW. We sweep the frequency of each MW component and calculate the purity from the retrieved single photon signals by coupling $\ket{91S_{1/2}, m_J=1/2}$ with $\ket{90 P_{3/2}}$. The H antenna is orthogonal with the quantum axis, and the purity is measured to be 82\%. The $\sigma_-$ antenna is $\sim 20^\circ$ relative to the quantum axis, and the purity is measured to be 56\%. The purity is far from unity, but because the middle state detuning is pretty large and there is no other disturbing states in 91$S_{1/2}$ except $\ket{r_1}$ and $\ket{r_2}$, we finally get high Raman fidelity. The unitary transformation is performed with another MW split from the preparation MW, with the pulse duration adjusted with another MW switch and the relative phase adjusted with a mechanical MW delay line.

\end{document}